\documentclass[aps,a4paper,showpacs,twocolumn,prl,floatfix]{revtex4}

\usepackage{graphicx}
\usepackage[ansinew]{inputenc}
\usepackage{array}
\usepackage{color}
\usepackage{amsmath}
\usepackage{amsxtra}
\usepackage{amstext}
\usepackage{amssymb}
\usepackage{latexsym}
\usepackage{dsfont}
\usepackage{float}

\usepackage{eucal}
\begin{document}

\title{Operators for the Aharonov-Anandan and Samuel-Bhandari Phases}

\author{P.-L. Giscard\footnote[2]{giscard@email.arizona.edu}}
\affiliation{Department of Physics and College of Optical
Sciences\\The University of Arizona, Tucson, Arizona 85721}

\date{\today}

\begin{abstract}
We construct an operator for the Aharonov-Anandan phase for time independent Hamiltonians. This operator is shown to generate the motion of cyclic quantum systems through an equation of evolution involving only geometric quantities, i.e. the distance between quantum states, the geometric phase and the total length of evolution. From this equation, we derive an operator for the Samuel and Bhandari phase (SB-phase) for non cyclic evolutions. Finally we show how the SB-phase can be used to construct an operator corresponding to a quantum clock which commutator with the Hamiltonian has a canonical expectation value.
\end{abstract}


\maketitle
Although experiments related to the geometric phase are known since Pancharatnam's work \cite{PanchaWork}, the concept of geometric phase was first recognized by M.V. Berry \cite{Berry1984}. In his seminal paper, Berry showed that an eigenstate of a parameter-dependent Hamiltonian varying adiabatically and cyclically, acquires both the well known dynamical phase and a gauge invariant phase that depends only on the geometry of the path taken by the Hamiltonian in the space of parameters. The mathematical explanation to this geometric phase was soon given by Simon \cite{Simon1983} whereas a generalization to Berry's phase was discovered by Aharonov and Anandan a few years later \cite{AAPhase1987}. The Aharonov-Anandan geometric phase (AA-phase) arises when a quantum system undergoes cyclic evolutions with time. The concept of geometric phase has since then been generalized to non cyclic, non unitary evolutions by Samuel and Bhandari \cite{Samuel1988} (SB-phase). However the condition of cyclicality required by the AA-phase, highly constrain the values of the geometric phase itself, which eventually led to a new algebraic method for the calculation of the AA-phase \cite{Giscard2008}. 

One of the fundamental axioms of quantum mechanics is that to any measurable physical quantity, an observable, corresponds an operator. Beyond the problems currently encountered with time-related quantities for which the construction of the corresponding operator(s) is still an unsolved question, researchers have produced operators for all other physical quantities. The AA-phase, as well as the SB-phase have been observed and measured experimentally in a large variety of systems (\cite{GeoPhaseWilczekShapere} and references therein), thus confirming their status of observables. Accordingly, there should exist operators for these quantities which construction is the main goal of this paper. We start by deriving a new expression for the AA-phase based on a preceding work by the same author \cite{Giscard2008}. This expression, which appears as an expectation value, is then used to construct the corresponding operator and to derive a selection rule giving the allowed and forbidden geometric phases depending on the state and the Hamiltonian evolving it. In a second part, the role of the operator for the AA-phase is explored through an new equation of motion for cyclic quantum systems which involves neither time nor any fundamental constant and from which we obtain the operator for the SB-phase. We continue by deriving an operator for the distance between quantum states as measured by the Fubini-Study metric along the curves solutions to Schr\"{o}dinger equation and the corresponding time-operator. We finish on some properties of the time-operator and show how it is related to a quantum clock.

The first step is the derivation of a new expression for the geometric AA-phase. We start from the results found in \cite{Giscard2008}. For a time independent Hamiltonian $\hat{H}$, provided that a state $|\psi\rangle$ undergoes a cyclic motion, the total AA-phase it acquires after one complete period of evolution is given by
\begin{equation}
\label{eq:aaphase}
\gamma=\phi+\frac{\tau_{\psi}}{\hbar}\langle\psi|\hat{H}|\psi\rangle,
\end{equation}
$\phi$ and $\tau_{\psi}$ being respectively the total phase and the period of the cyclic motion. These quantities depend on both the Hamiltonian $\hat{H}$ and the state $|\psi\rangle$. Let $B=\{|\phi_{k}\rangle\}$ be a basis in which $\hat{H}$ is diagonal and $\Lambda=\{\lambda_{k}\}$ the corresponding set of its eigenvalues. 
Let $B_{\psi}\subseteq B$ be the smallest set of eigenvectors needed to decompose the state $|\psi\rangle$ on $B$ and let $\Lambda_{\psi}\subseteq\Lambda$ be the corresponding set of eigenvalues. Finally, let $\Delta E_{\psi}$ be the set of non-zero energy spacings in $\Lambda_{\psi}$, i.e. $\Delta E_{\psi}=\{\Delta E_{k,i}=\lambda_{k}-\lambda_{i}\}_{\lambda_{k,i}\in\Lambda_{\psi},~\lambda_{k}\neq\lambda_{i}}$. Then from \cite{Giscard2008} we have
\begin{eqnarray}
\tau_{\psi}&=&2\pi\hbar ~\mathrm{LCM}\left(\Delta E_{\psi}^{-1}\right),\label{eqn:tau}\\
\phi&=&2\pi\left[m-\lambda~\mathrm{LCM}\left(\Delta E_{\psi}^{-1}\right)\right],\label{eqn:phi}
\end{eqnarray}
where LCM means least common multiple and $m$ is an integer depending on $\lambda$. Note that Eq.(\ref{eqn:phi}) is valid for any $\lambda\in\Lambda_{\psi}$. For the sake of clarity, we will use $\mathrm{L}_{\psi}=\mathrm{LCM}\left(\Delta E_{\psi}^{-1}\right)$ in the following. 

To obtain an expression for the geometric phase $\gamma$, we combine Eq.(\ref{eq:aaphase}), Eq.(\ref{eqn:tau}) and Eq.(\ref{eqn:phi}). The expression for the total phase $\phi$ is calculated using an arbitrary $\lambda_{j}\in\Lambda_{\psi}$. Moreover, the expectation value $\langle\psi|\hat{H}|\psi\rangle$ is evaluated on $B$ to be
\begin{equation}
\langle\psi|\hat{H}|\psi\rangle=\sum_{\lambda_{i}\in\Lambda_{\psi}}\lambda_{i}|\phi_{i}|^{2},
\end{equation}
where $\phi_{i}=\langle\phi_{i}|\psi\rangle$. By combining these expressions we obtain the following expression of the total AA-phase for a normalized state $|\psi\rangle$ undergoing cyclic evolutions
\begin{equation}
\label{eq:aaphasenew}
\gamma_{\psi}=2\pi m_{j}+2\pi\sum_{\lambda_{i}\in\Lambda_{\psi}}(\lambda_{i}-\lambda_{j})\mathrm{L}_{\psi}|\phi_{i}|^2,
\end{equation} 
$m_{j}$ being an integer that depends on $\lambda_{j}$. Let us now define $p_{\psi, ij}=(\lambda_{i}-\lambda_{j})\mathrm{L}_{\psi}$. Independently of the state under consideration, the coefficients $p_{\psi, ij}$ have the following properties : 
\begin{eqnarray}
\label{eq:propertiesp}
p_{ij}&=&-p_{ji},\label{eqn:propp1}\\
p_{ij}&=&p_{kj}-p_{ki},\label{eqn:propp1}\\
p_{ij}&=&m_{i}-m_{j}.\label{eqn:propp1}
\end{eqnarray}
The last equality is obtained by equalizing Eq.(\ref{eq:aaphasenew}) calculated with two different $\lambda_{j}$. From the above properties, we obtain a new expression for the total AA-phase accumulated by the state $|\psi\rangle$ after one cyclic evolution
\begin{eqnarray}
\gamma_{\psi}&=&2\pi\sum_{i}p_{\psi, ij}|\phi_{i}|^{2}~[2\pi],\label{eqn:gammap}\\
\gamma_{\psi}&=&2\pi\sum_{i}m_{i}|\phi_{i}|^{2}~[2\pi],\label{eqn:gammam}
\end{eqnarray}
with both $m$ and $p$ integers. Furthermore, the above formulas Eq.(\ref{eqn:gammap}) and Eq.(\ref{eqn:gammam}) are valid in the diagonal basis of the considered Hamiltonian $\hat{H}$ and therefore, cannot be simultaneously correct for two non commuting Hamiltonians $\hat{H}$ and $\hat{H'}$.  

Let us now turn to the problem of finding an operator for the total AA-phase. Interestingly, Eq.(\ref{eqn:gammap}) (and equivalently Eq.(\ref{eqn:gammam})) can be used to define such an operator, that we call the geometric-operator $\hat{G}_{\psi}$ (the notation is explained in the following), which expectation value for states undergoing cyclic evolutions when subjected to some time independent Hamiltonian is the total AA-phase 
\begin{equation}
\label{eq:expectG}
\gamma_{\psi}=\langle \hat{G}_{\psi}\rangle.
\end{equation}
From Eq.(\ref{eqn:gammap}), the expression of $\hat{G}_{\psi}$ in the basis $B$ is given by
\begin{equation}
G_{\psi,kl}=2\pi p_{\psi, kj} \delta_{k,l},
\end{equation}
or equivalently
\begin{equation}
\label{eq:stateopgeo}
\hat{G}_{\psi}=\sum_{i}2\pi p_{\psi, ij} |\phi_{i}\rangle\langle\phi_{i}|.
\end{equation}
From Eq.(\ref{eq:aaphase}) and Eq.(\ref{eqn:phi}), the relation to the Hamiltonian is the following 
\begin{equation}
\label{eq:GfunctionH1}
\hat{G}_{\psi}=\frac{\tau_{\psi}}{\hbar}(\hat{H}-\lambda_{j}),
\end{equation}
with $\lambda_{j}\in\Lambda_{\psi}$. This operator depends on the considered state $|\psi\rangle$ as both $\tau_{\psi}$ and $\Lambda_{\psi}$ depend on $|\psi\rangle$. To emphasize this state-dependence, we denote $\hat{G}_{\psi}$ the geometric operator for state $|\psi\rangle$. This operator is necessarily the same for any state $|\varphi\rangle$ located on the path $C$ solution to Schr\"{o}dinger equation passing through $|\psi\rangle$ and could equivalently be denoted $\hat{G}_{\varphi\in C}$. Thus, although $\hat{G}_{\psi}$ is linear once $|\psi\rangle$ has been specified, an operator acting on the whole space of states and with expectation value the AA-phase is non-linear and accepts no matrix representation. Only the state dependent form $\hat{G}_{\psi}$ accepts one (given in Eq.(\ref{eq:stateopgeo})). 

Before studying the role of the geometric-operator in the motion of cyclic quantum systems, let us consider some consequences of Eq.(\ref{eqn:gammap}). First, note that a simplification of Eq.(\ref{eqn:gammap}) occur for a two-level states. As shown in \cite{Giscard2008} two-level quantum states (i.e. $\Lambda_{\psi}$ contains exactly two different elements) always undergo cyclic evolutions, whatever the time independent Hamiltonian considered. Using the definition of the $p_{i}^{j}$ coefficients of Eq.(\ref{eqn:gammap}), we see that $p_{0}^{1}=-1$ if $\lambda_{0}<\lambda_{1}$ and $p_{0}^{1}=1$ otherwise. This observation leads to 
\begin{equation}
\label{eq:twolevels}
\gamma=-2\pi|\phi_{0}|^2=2\pi|\phi_{1}|^2~[2\pi],
\end{equation}
if $\lambda_{0}<\lambda_{1}$ and all the signs are reversed otherwise. Beyond these signs, the only Hamiltonian dependence of this equation is contained in the $|\phi_{i}|^{2}$. This is consistent with the fact that the dynamics of any two-level system can be mapped on the dynamics of a spin-$1/2$ particle in a fictitious magnetic field. Indeed this consideration entail that, given that we work in the eigenbasis of the Hamiltonian, the geometric phase for a spin-$1/2$ particle is the same than for any other two-level Hamiltonian (up to a sign). The geometric-operator $\hat{G}_{\psi}$ for $|\psi\rangle=e^{i\Phi}\mathrm{cos}(\theta/2)|\phi_{0}\rangle+\mathrm{sin}(\theta/2)|\phi_{1}\rangle$ takes a particularly simple form 
\begin{equation}
\label{eq:Gmatrix2level} \hat{G}_{\psi} =
\begin{pmatrix}
\pm2\pi &  0\\
0 & 0  
\end{pmatrix}.
\end{equation} 
Then, Eq.(\ref{eq:twolevels}) or Eq.(\ref{eq:expectG}, \ref{eq:Gmatrix2level}), immediately give the well known result 
\begin{equation}
\gamma=\pm\pi(1-\mathrm{cos}(\theta))[2\pi].
\end{equation}

A second notable consequence of Eq.(\ref{eqn:gammap}) gives a rule of selection for geometric-phases. Let $\Xi$ be a set of time-independent Hamiltonians that all commute with one another and $|\psi\rangle$ a normalized state that can be expressed in the common eigenbasis $B$ of $\Xi$ as
\begin{equation}
\label{eq:formomega}
|\psi\rangle=\frac{1}{\sqrt{\omega}}\sum_{|\phi_{i}\rangle\in B_{\psi}}\alpha_{i}e^{i\theta_{i}}|\phi_{i}\rangle,
\end{equation}
with all $\alpha_{i}$ integers, $\theta_{i}$ real numbers, and $\omega$ is the smallest positive \textit{integer} satisfying Eq.(\ref{eq:formomega}). Note that such a form exists if and only if all $|\langle\phi_{i}|\Psi\rangle|^{2}_{i\in[0,N]}$ are rational numbers. Thus we will say that states obeying Eq.(\ref{eq:formomega}) accept a rational decomposition on $B$. From the fact that the $p_{i}^{j}$ coefficients of Eq.(\ref{eqn:gammap}) are integers, it follows that states accepting a rational decomposition on $B$, can only acquire AA-phases of the form 
\begin{equation}
\gamma_{\psi}=\frac{2n\pi}{\omega}[2\pi],~n~\mathrm{integer},
\end{equation}
whatever the Hamiltonian(s) of $\Xi$ and/or the number of consecutive cyclic evolutions considered.
This shows that \textit{only} the geometric phases that are \textit{rational multiples of $\pi$} are allowed for state accepting a rational decomposition on $B$. This is an intrinsic property of both the state $|\psi\rangle$ and the set $\Xi$. Note that the rationality of $\gamma_{\psi}/\pi$ implies a finite number of cyclic evolution giving different geometric phases. At the opposite, a state $|\varphi\rangle$ that does not accept a rational decomposition on $B$ can \textit{only} acquire AA-phases that are \textit{irrational multiples of $\pi$} when subjected to cyclic evolution(s) by some Hamiltonian(s) of $\Xi$. Interestingly this implies in principle the possibility of distinguishing any number of cyclic evolutions from any other. Indeed since $\gamma_{\varphi}/\pi$ is irrational, $(p\gamma_{\varphi})[2\pi]\neq (q\gamma_{\varphi})[2\pi]$, for any integers $p\neq q$ representing the number of cyclic evolutions elapsed. However, it can be shown that there will always be a number $q'$ of cyclic evolutions so that the resulting geometric phase is arbitrarily close to $\gamma_{\varphi}[2\pi]$ which means that distinguishing an infinite number of cyclic evolutions would experimentally require an infinite precision in the measurement of the geometric-phase. 

Let us now focus on the role of the geometric operator in an equation of motion for cyclic quantum systems. To know how evolves the state of a quantum system, a parameter evolving along the path $C$ solution to Schr\"{o}dinger equation in the Hilbert space is required. This parameter is generally chosen to be the time. But in a paper about the evolution of quantum systems \cite{Anandan1990}, J. Anandan and Y. Aharonov, gave the expression of the instantaneous speed of evolution of a quantum system $|\psi(t)\rangle$ in the space of quantum states as 
\begin{equation}
\label{eq:speedevo}
\vartheta_\psi(t)=\frac{\Delta \hat{H} _{\psi}(t)}{\hbar},
\end{equation}
where $\Delta \hat{H}_{\psi}(t)=(\langle\psi(t)|\hat{H}^{2}|\psi\rangle-\langle\psi(t)|\hat{H}|\psi(t)\rangle^2)^{1/2}$ is the uncertainty in energy of the state $|\psi(t)\rangle$. For time independent Hamiltonians, this speed in a constant of motion and the passage from a time parametrization to a distance parametrization of the Schro\"{o}dinger equation is straightforward. In \cite{Anandan1990}, J. Anandan and Y. Aharonov, showed indeed that the distance $s$ as measured by the Fubini-Study metric in the space of quantum states since the beginning of the unitary evolution along a curve $C$ is given by
\begin{equation}
\label{eq: dists}
s=\int_{C}{\frac{\Delta \hat{H} (t)_{\psi\in C}}{\hbar}dt}\equiv\int_{C}\vartheta_{\psi}(t)dt,
\end{equation}
which reduces to $s=\frac{\Delta \hat{H} _{\psi\in C}}{\hbar}t$ for time independent Hamiltonians, i.e. the time elapsed times the speed of evolution. This allows to write a simple equation of evolution equivalent to the Schr\"{o}dinger equation but using the distance $s$ as parameter
\begin{equation}
\label{eq: schrointer}
i\hbar\frac{\partial |\psi\rangle}{\partial s} \frac{ds}{dt}=\hat{H}|\psi\rangle.
\end{equation}
Eq.(\ref{eq: dists}) gives $\frac{ds}{dt}=\frac{\Delta \hat{H}_{\psi}}{\hbar}$ and, for states undergoing cyclic evolution, we can use Eq.(\ref{eq:GfunctionH1}) to get $\Delta \hat{G}_{\psi}=\tau_{\psi} \Delta \hat{H}/\hbar$. These manipulations lead to
\begin{equation}
\label{eq:eqevogeo}
i\Delta \hat{G}_{\psi} \frac{\partial |\psi(s)\rangle}{\partial s}=(\hat{G}_{\psi}+\frac{\lambda\tau_{\psi}}{\hbar})|\psi(s)\rangle.
\end{equation}
which yields, by dropping the $\lambda\tau_{\psi}/\hbar$ constant term \cite{gaugechange},
\begin{equation}
\label{eq:eqevogeo1}
i\Delta \hat{G}_{\psi} \frac{\partial |\psi(s)\rangle}{\partial s}=\hat{G}_{\psi}|\psi(s)\rangle.
\end{equation}
Now since the evolution is cyclic in time, it is also cyclic in distance, i.e. there exists a period in length $S_{\psi}$ such that, $n$ being an integer, $|\psi(nS_{\psi}+s)\rangle=|\psi(s)\rangle$ upto a phase factor. Logically $S_{\psi}= \tau_{\psi} \vartheta_{\psi}$ thus $S_{\psi}\equiv\Delta G_{\psi}$ and the motion of a cyclic quantum system is govern by
\begin{eqnarray}
iS_{\psi}\frac{\partial |\psi(s)\rangle}{\partial s}=\hat{G}_{\psi}|\psi(s)\rangle,\label{eqn:eqevogeos}\\
i\tau_{\psi}\frac{\partial |\psi(t)\rangle}{\partial t}=\hat{G}_{\psi}|\psi(t)\rangle\label{eqn:eqevogeot}.
\end{eqnarray}
in the two parametrizations. These two equations are valid for any quantum state undergoing cyclic evolutions and only for them, or in other terms, their solutions are the cyclic solutions to Schr\"{o}dinger equation. Note that Eq.(\ref{eqn:eqevogeos}) can be considered as a purely geometric equation of motion as it involves only geometric objects, the geometric-operator (or equivalently the geometric-phase), the distance between quantum state as measured by the Fubini-Study metric and the total length of a cyclic evolution. Now we can use Eq.(\ref{eqn:eqevogeos}) to get another expression for the AA-phase as
\begin{equation}
\gamma_{\psi}=iS_{\psi} \langle\frac{\partial}{\partial s}\rangle,
\end{equation}
and a new expression for the geometric-operator for state $|\psi\rangle$ as 
\begin{equation}
\label{eq:opegeonew}
\hat{G}_{\psi}=iS_{\psi}\frac{\partial}{\partial s}.
\end{equation}
A simple example of a solution to Eq.(\ref{eqn:eqevogeos}) is given by a two level system. The geometric operator is given by Eq.(\ref{eq:Gmatrix2level}) and let the initial state be $|\psi\rangle=e^{i\Phi}\mathrm{cos}(\theta/2)|\phi_{0}\rangle+\mathrm{sin}(\theta/2)|\phi_{1}\rangle$. The state at distance $s$ is then 
\begin{equation}
|\psi(s)\rangle=e^{\pm i\frac{s}{\vert\phi_{0}\vert\vert\phi_{1}\vert}}e^{i\Phi}\mathrm{cos}(\theta/2)|\phi_{0}\rangle+\mathrm{sin}(\theta/2)|\phi_{1}\rangle.
\end{equation}
where we have used $S_{\psi}=\Delta \hat{G}_{\psi}=2\pi|\phi_{0}||\phi_{1}|$. The only Hamiltonian dependence is contained in the decomposition of the initial state on the eigenbasis $\{|\phi_{0}\rangle$, $|\phi_{1}\rangle\}$. Again, this is consistent with the possibility of mapping the dynamics of any two level systems on the one of a spin-1/2 in a magnetic field.

To find an operator for the SB-phase, we now introduce a third parametrization of the Schr\"{o}dinger equation based on the instantaneous geometric phase $\gamma(s)$, which is the geometric phase accumulated at a given distance $s$. This quantity can be calculated following the work by Samuel and Bhandari \cite{Samuel1988} who generalized the notion of geometric phase to non necessarily closed evolutions of quantum systems. It is found to be simply $\gamma(s)=s\gamma_{\psi}/S_{\psi}$. This together with Eq.(\ref{eq:opegeonew}), lead us to propose the instantaneous geometric-operator $\hat{G}(s)$
\begin{equation}
\label{eq:opegeoinst}
\hat{G}(s)=is\frac{\partial}{\partial s},
\end{equation}
which is state independent. This is a generalization of the operator for the AA-phase and corresponds to the SB-phase in the case of time independent Hamiltonians. In this parametrization we can contruct an operator $\hat{S}$ for the distance elapsed since an arbitrary point $|\psi(s=0)\rangle$ and so that $s=\langle\psi(s)|\hat{S}|\psi(s)\rangle$. Indeed, since the instantaneous geometric phase fulfill $\gamma(s)=s\gamma_{\psi}/S_{\psi}$. Thus $s=\gamma(s)S_{\psi}/\gamma_{\psi}$ and we propose
\begin{equation}
\hat{S}(s)=\hat{G}(s)\frac{S_{\psi}}{\gamma_{\psi}}.
\end{equation}
Upon using Eq.(\ref{eq:opegeoinst}) and remarking that $S_{\psi}/\gamma_{\psi}=\partial s/\partial \gamma(s)$, we obtain
\begin{equation}
\hat{S}(s)=is\frac{\partial}{\partial \gamma(s)}.
\end{equation}
This is consistent with the fact that $\langle i\frac{\partial}{\partial \gamma(s)}\rangle=1$, which can be demonstrated using Eq.(\ref{eq:opegeoinst}). It is now straightforward to obtain a similar expression for a time-operator $\hat{T}$ with the parameter $t$ entering Schr\"{o}dinger equation as expectation value. Indeed using the speed of quantum evolution we have $\hat{T}(s)=\hat{S}(s)/\vartheta_{\psi}$, which gives on using $s/\vartheta_{\psi}=t$ and $s=\gamma(s)S_{\psi}/\gamma_{\psi}$,
\begin{eqnarray}
\hat{T}(t)&=&it\frac{\partial}{\partial \gamma(\vartheta_{\psi}t)},\label{eqn:expre1T}\\ 
\hat{T}(s)&=&is\frac{\tau_{\psi}}{\gamma_{\psi}}\frac{\partial}{\partial s}.\label{eqn:expre2T}
\end{eqnarray}
Note that, since the geometric-operator is Hermitian and since $S_{\psi}$ is real, Eq.(\ref{eq:opegeonew})  leads to $(i\partial/\partial s)^{\dagger}=i\partial/\partial s$. Therefore, $s$, $\tau_{\psi}$ and $\gamma_{\psi}$ being real, we conclude using Eq.(\ref{eqn:expre2T}) that $\hat{T}$ is Hermitian. Interestingly the commutators of $\hat{T}$ with $\hat{H}$ and $\hat{G}_{\psi}$ can be calculated and are found to be
\begin{eqnarray}
\lbrack\hat{H},\hat{T}(s)\rbrack&=&-\hbar\frac{\partial}{\partial \gamma(s)},\\
\lbrack\hat{G}_{\psi},\hat{T}(s)\rbrack&=&-\tau_{\psi}\frac{\partial}{\partial \gamma(s)},
\end{eqnarray}
with, as $\langle i\frac{\partial}{\partial \gamma(s)}\rangle=1$, the following expectation values 
\begin{eqnarray}
\langle\lbrack\hat{H},\hat{T}(s)\rbrack\rangle&=&i\hbar,\\
\langle\lbrack\hat{G}_{\psi},\hat{T}(s)\rbrack\rangle&=&i\tau_{\psi},
\end{eqnarray}
remarkably the first quantity is both time and state independent. Furthermore, at $t=\tau_{\psi}$, the operator $\hat{T}(t)$ becomes $\hat{T}(\tau_{\psi})=i\tau_{\psi}\frac{\partial}{\partial \gamma_{\psi}}$ which, on using the relation between geometric phase $\gamma_{\psi}$ and energy $\epsilon=\langle \hat{H}\rangle-\lambda$, $\tau_{\psi}\epsilon=\hbar \gamma_{\psi}$ (see Eq.(\ref{eq:GfunctionH1})) leads to
\begin{equation}
\hat{T}(\tau_{\psi})=i\hbar\frac{\partial}{\partial \epsilon}.
\end{equation}
This operator has already been proposed for time in quantum mechanics, see \cite{Holevo1982}. 

We emphasize that the operators $\hat{S}$ and $\hat{T}$ proposed here correspond respectively to the distance and the time elapsed since an arbitrarily chosen initial state. In other terms if the expectation value of $\hat{T}$ was obtained on a ensemble of quantum systems at two different moments, then the difference between those expectation values would be the time elapsed between the two series of measurements. This could therefore be used as quantum clock measuring the time of unitary Schr\"{o}dinger-evolution. 

\begin{acknowledgments}
We thank M. Bhattacharya for starting our interest in geometric phases and for helpful discussions and  P. Meystre whose support allowed this work.
\end{acknowledgments}

\bibliography{AALetter}

\begin{thebibliography}{10}
\bibitem[{gau()}]{gaugechange}
\bibinfo{note}{This gauge change does not change the geometric-phase nor the
  dynamics of the system.}
\end{thebibliography}
\end{document}